\documentstyle[preprint,aps]{revtex}

\begin{document}
\title{Experimental application of decoherence-free subspaces in a
quantum-computing algorithm}
\author{Masoud Mohseni, Jeffrey S. Lundeen, Kevin J. Resch, and Aephraim M.
Steinberg }
\address{Department of Physics, University of Toronto, 60 St. George St.,\\
Toronto, Ontario, Canada, M5S 1A7}
\date{20 Dec 2002}
\maketitle

\begin{abstract}
For a practical quantum computer to operate, it will be essential to
properly manage decoherence. One important technique for doing this is the
use of ``decoherence-free subspaces''\ (DFSs), which have recently been
demonstrated. Here we present the first use of DFSs to improve the
performance of a quantum algorithm. An optical implementation of the
Deutsch-Jozsa algorithm can be made insensitive to a particular class of
phase noise by encoding information in the appropriate subspaces; we observe
a reduction of the error rate from 35\% to essentially its pre-noise value
of 8\%.
\end{abstract}

One of the great stumbling blocks to building quantum computers, with their
oft-touted ability to resolve certain problems more efficiently than any
classical algorithm \cite{Feynman,nchuang} is the ubiquity of decoherence.
Coupling of any element of a quantum computer to an environment destroys its
unitary evolution, and introduces uncontrollable noise; at first, it was
thought by many \cite{Landauer} that these errors would make quantum
computation impossible in practice. Since then, a variety of techniques for
correcting errors and/or building in immunity to certain classes of
decoherence have been developed \cite{QECc,DFS,Lidar} and it has been proved
that if errors are kept below a certain constant threshold, arbitrarily
large quantum computers are possible\cite{thresholdKnill}. One important
technique involves computing within certain subspaces of the full system's
Hilbert space known as decoherence-free subspaces (DFSs) \cite{DFS,Lidar}
which remain unaffected by the interaction with the environment. Such DFSs
exist when the interaction Hamiltonian has an appropriate symmetry property.
DFSs have been demonstrated in a linear-optical experiment \cite{kwiatDFS}
and in NMR \cite{NMRDFS} and recently to help circumvent the technical noise
which had previously limited the operation of ion-trap quantum computers %
\cite{iontDFS}. To date, no demonstration has been made of the usefulness of
DFSs in the context of the implementation of an actual quantum-computing
algorithm\footnote{%
Just prior to submission of this manuscript, we learned that a similar
demonstration has now been perfoermed in liquid-state NMR: J. Ollerenshaw,
D. Lidar, and L. Kay, in preparation.}. In this paper, we present a
linear-optical implementation of the two-qubit Deutsch-Jozsa algorithm \cite%
{DJ92,nchuang}, and demonstrate that when a certain class of noise is
introduced into the system, greatly increasing the error rate of the
algorithm, it is possible to `encode' one logical qubit into two physical
qubits and take advantage of DFSs, reducing the error rate to close to zero.

Linear optics is well known to be an extremely powerful arena for the
transportation and manipulation of quantum information, \cite%
{BenBrassard,ReckZeilinger}. Although it is also well known that due to the
linearity of optics, this arena does not allow for scalable construction of
quantum gates \cite{Cerf}, recent work has shown that the incorporation of
detection and post-selection may in fact render all-optical quantum
computers an attractive possibility \cite{KLM}. Work also proceeds on
development of nonlinearities which would allow for the development of
natural two-qubit gates in optics\cite{Kimble}. While we do not yet have
access to a truly scalable optical quantum-computer architecture, many of
the elements of any such system would be identical to those used in simple
linear-optical geometries \cite{Cerf}. For this reason, linear optics
remains an important domain for the study of quantum coherence and error
correction, even while the ultimate fate of optical quantum computing is
uncertain. Recently, striking demonstrations of quantum search algorithms %
\cite{Spreeuw} have been carried out in linear-optical systems, as has the
first verification of DFSs \cite{kwiatDFS}. Additionally, it is already
clear that even if quantum computation never becomes truly practical,
quantum information processing may have a great effect on the practice of
communications and cryptography\cite{TeleportZeilinger,Gisin}. Although some
information-processing will be necessary in this area as well, the question
of scalability is not crucial, and linear-optical quantum computation could
well prove applicable for elements such as quantum repeaters \cite%
{RepeatersBriegel}. In this context, we have chosen to study the
applicability of DFSs to a linear-optical implementation of the quantum
Deutsch-Jozsa algorithm, despite the non-scalable nature of the present
architecture.

The Deutsch-Jozsa algorithm is designed to distinguish between two classes
of functions (``oracles'')\ on N-bit binary inputs. ``Constant''\ functions
return the same value ( 0 or 1) for all $2^{n}$ possible inputs, while
``balanced''\ functions return 0 for half the possible inputs and 1 for the
other half. \ Clearly, a classical algorithm would on some occasions require
as many as $2^{n-1}+1$ queries to unambiguously determine to which class a
given oracle belongs. \ By contrast, Deutsch and Jozsa showed \cite{DJ92}
that a quantum algorithm requires only one such query. \ In the 2-qubit
Deutsch-Jozsa algorithm \cite{nchuang}, the oracle is a function on a single
bit. \ It takes as input a query bit $x$ and a signal bit $y$; its action is
to perform the unitary mapping $\left| x,y\right\rangle ${\it \ }$%
\rightarrow ${\it \ }$\left| x,y\oplus f(x)\right\rangle .$ To perform the
algorithm, the input is prepared in $H\left| 0\right\rangle \otimes H\left|
1\right\rangle =\frac{1}{2}[\left| 0\right\rangle +\left| 1\right\rangle
]\otimes \lbrack \left| 0\right\rangle -\left| 1\right\rangle ]$ which the
oracle maps to $\frac{1}{2}[\left| 0\right\rangle \otimes (\left|
f(0)\right\rangle -\left| \overline{f(0)}\right\rangle )+\left|
1\right\rangle \otimes (\left| f(1)\right\rangle -\left| \overline{f(1)}%
\right\rangle ]=\frac{1}{2}[\left| 0\right\rangle e^{i\pi f(0)}+\left|
1\right\rangle e^{i\pi f(1)}]\otimes H\left| 1\right\rangle $. \ A Hadamard
on the query qubit then transforms it into $\left| f(0)\oplus
f(1)\right\rangle ,$ which is equal to 0 for constant and 1 for balanced
functions. Thus measurement in the computational basis allows one to
determine a global property of $f(x)$ , namely $f(0)\oplus f(1),$ in a
single evaluation of the function. Furthermore, the signal qubit is in fact
superfluous after the oracle \cite{CollinsRed}. \ Thus only one logical
qubit is needed after the operation of the oracle. \ If some source of
decoherence is present during the propagation from the oracle to the final
Hadamard, one may consider encoding this logical qubit in some
decoherence-free subspace of the two physical qubits.

In this experiment we represent the four basis states of two logical qubits (%
$\left| 00\right\rangle ,\left| 01\right\rangle ,\left| 10\right\rangle $
and $\left| 11\right\rangle $, where the first bit corresponds to the query
and the second to the signal) by a photon traveling down one of four optical
rails numbered 1, 2, 3 and 4 respectively. It is possible to implement a
universal set of one- and two-qubit operations in a four-rail representation %
\cite{Cerf}. For example a NOT gate on the query qubit can be realized by
simultaneously swapping rails 1 and 3 and rails 2 and 4. A CNOT gate on the
signal qubit is implemented by swapping rails 3 and 4. \ To perform a
Hadamard gate on the query qubit, we combine rails 1 and 3 and rails 2 and 4
at two 50/50 beam-splitters; a $\pi $ phase shift is also needed on two of
the arms.Analogous gates can be constructed for the other qubit. The
transformations introduced by the four possible functions can also be
implemented in this representation by four different settings of an oracle
operating as follows: if $f(0)$ is 1, rails 1 and 2 are swapped; if $f(1)$
is 1, rails 3 and 4 are swapped. \ Thus the task of distinguishing balanced
from constant oracles reduces to that of determining whether the number of
swaps was odd or even.

The schematic diagram of the interferometer is shown in Fig. 1. Each photon
is sent along rail 2 corresponding to the logical state $\left|
01\right\rangle $. The two pairs of 50/50 beam splitters $A1$, $A2_{\text{ }%
} $and $B1$, $B2$ implement the two Hadamard gates on the query and signal
qubits respectively, preparing the qubits for the oracle's action. The last
two 50/50 beam splitters $C1$ and $C2$ realize the Hadamard gate on the
query qubit after the oracle. Rails 1-4 illuminate photodiodes $PD1-PD4$. \
A photon reaching PD1 or PD2 indicates that the value of the query qubit
after the algorithm, $f(0)\oplus f(1)$ , is 0. \ This constitutes a
determination that the oracle is constant, while the other two detectors
indicate balanced oracles.

One source of decoherence in such systems is the phase noise introduced by
fluctuating optical path lengths between the different sections of the
apparatus, created either by variations in distance or by temperature
variations and turbulent air flow. In real optical systems, the stability of
certain path-length differences may be larger than that of others, either
because of the physical proximity of certain paths to one another or because
of the particular sources of mechanical or thermal noise. This may lead to a
situation where the dominant source of decoherence exhibits a particular
symmetry which can be exploited for computing within DFSs. To simulate the
effects such processes could have in larger-scale, distributed
quantum-information systems, we introduced a high degree of turbulence by
placing the tip of a hot soldering iron below two of the optical rails.
These two optical paths (rails 2 and 3) were spatially superposed in this
region, distinguished only by their polarisation; for this reason, they
experienced essentially the same random phase shifts under the influence of
the turbulent air flow, relative to the other optical rails. Since the
outputs of the optical Deutsch-Jozsa setup are the outputs of two parallel
interferometers, which measure the phase of rail 2 with respect to that of
rail 4 and rail 3 with respect to rail 1, this phase noise destroys the
interference on which the success of the algorithm relies. On the other
hand, inspection of the optical schematic makes the physical process behind
the algorithm evident: rails 1 and 3 are prepared in phase with one another,
while rails 2 and 4 are also prepared in phase, but $180^{\circ }$out of
phase with the former pair. Thus, constructive interference is observed
either between 1 and 3 or between 2 and 4. If a single pair (1 and 2 or 3
and 4) is swapped by the oracle, destructive interference is instead
observed at both interferometers, while if an even number of swaps occurs,
constructive interference is restored. In other words, so long as each
interferometer compares an output of each of the potential swap regions in
the oracle with one from the other, it is possible to distinguish a balanced
oracle (one swap) from a constant oracle (zero or two swaps). The strategy
to deal with phase noise impressed symmetrically on paths 2 and 3 now
becomes clear: instead of interfering 2 with 4 and 1 with 3, one can
accomplish the same task by interfering 2 with 3 and 1 with 4. In this way,
the random phase appears at both inputs to the same interferometer, and has
no effect on the measured results.

This modification can be expressed as an encoding of the data into a pair of
DFSs. Since our engineered phase noise has identical effects on the two
states of odd parity ($\left| 01\right\rangle $ and $\left| 10\right\rangle
, $ stored on rails 2 and 3 respectively) and on the two states of even
parity ($\left| 00\right\rangle $ and $\left| 11\right\rangle $, stored on
rails 1 and 4), each fixed-parity subspace can store a single logical qubit
in a decoherence-free fashion. The action of the soldering iron tip may be
modelled by the evolution operator $\exp (i\sigma _{z}^{1}\sigma
_{z}^{2}\delta \phi )$, where $\delta \phi $ is a random, fluctuating phase.
In a subspace with a definite eigenvalue of $\ \sigma _{z}^{1}\sigma
_{z}^{2} $, the random phase, $\delta \phi ,$ does nothing but impress an
overall global phase on the quantum state, leaving the information within
the subspace unaffected. Since the 2-qubit Deutsch-Jozsa algorithm relies on
a single qubit (query qubit) after the oracle has completed its action, this
single qubit may be encoded in either of these DFSs, providing immunity to
parity-dependent phase noise which occurs between the oracle and the final
Hadamard gates. \ As shown in Fig. 1c, a CNOT after the oracle encodes the
query qubit into these DFSs, and a second CNOT after the final Hadamard can
be used for decoding. \ The decoding CNOT is unnecessary since measurements
are only performed on the query qubit. \ Swapping rails 3 and 4 performs the
encoding; or equivalently, beam-splitters $C1$ and $C2$ may be replaced by $%
D1$ and $D2$.

The actual experimental setup is shown in Fig. 2. The light source was a
diode laser operating at 780 nm. To implement the four different oracle
settings a specific kind of variable beam splitter (VBS) was designed. This
variable beam splitter consists of a half-waveplate between two polarizing
beam splitters (PBS); any desired reflectivity can be obtained with this
optical arrangement. For realizing our oracles, a pair of these VBSs was
used, and each was adjusted either for maximum or minimum reflectivity,
essentially acting as a swap or the identity. \ Hadamards were constructed
using similar VBSs. After the oracle rails 2 and 3 were combined into the
same spatial mode in a PBS to guarantee the collective phase shift for these
beams in presence of decoherence, and then separated out by another PBS. The
transformation between two different encodings was performed by applying
another VBS to either swap rails 3 and 4 or not. The experimental setup was
designed such that in all of these interferometers the spatial path lengths
are always balanced. The average fringe visibility for all four output ports
and all possible settings of oracle and encoding was measured to be about
95\%. This setup consists of 16 different possible Mach-Zehnder
interferometers, of which two are in operation for any given oracle/encoding
combination.\ 

The experiment was performed by measuring the signals at detectors $PD1$
through $PD4$ as the half waveplates were adjusted to cycle through all four
oracles and both encodings. The intensities at detectors PD1 through PD4
were normalized to their sum, to yield the probabilities of a photon
reaching each of the detectors. \ These normalized intensities are plotted
in Fig. 3 for all 4 oracle settings, in both the standard algorithm and the
DFS encoding. \ Ideally, all the photons should arrive at detectors $PD1$
and $PD2$ for constant functions and at detectors $PD3$ and $PD4$ for
balanced functions. \ In figure 4, we plot the probability of a photon
reaching{\it \ either} PD1 or PD2, and the probability of a photon reaching 
{\it either} PD3 or PD4. \ The average error rates were measured to be about
8\% in the absence of added noise. The sources of errors in this experiment
were mostly due to imperfect visibility, (due to alignment and waveplate
setting), and uncertainty and drift in the optical phases setting, when a $%
12^{\circ }$ phase error on one beam correspond to a 2\% error rate. The
drift of the interferometer during measurement was kept low by balancing all
path lengths as mentioned above and enclosing the interferometer.
Introduction of the turbulent airflow increased the average error rates to
35\% for the standard algorithm. When the DFS encoding was used in the
presence of turbulence, however, the error rates dropped to 7\%, essentially
equal to the value in the absence of noise.

This work demonstrates that a simple modification of a quantum algorithm may
be used to encode information into DFSs, and significantly reduce the error
rate introduced by realistic, physical noise sources, provided that these
sources have certain symmetry properties. Phase noise is an everpresent
issue in coherent optical systems, and often exhibits certain correlations
which should be exploitable in this manner. We also note that since the
noise characteristics are intimately tied to the particular physical
realization of a quantum circuit, it may often prove easier to design the
decoherence-free process by direct consideration of the multiple
interferometers which constitute the optical device, than by contemplation
of the very general quantum circuits.

This work was supported by The US Air Force Office of Scientific Research
(F49620-01-1-0468), NSERC, and Photonics Research Ontario. We would like to
acknowledge useful discussions with Daniel Lidar and also thank Guillaume
Foucaud and Chris Ellenor for technical assistance.

\bigskip

\bigskip

\bigskip

\bigskip

\bigskip

\bigskip

\bigskip

\bigskip

\bigskip

\bigskip

\bigskip

\section{\protect\bigskip Figure Captions}

Fig 1. Schematic of an interferometer which implements the 2-qubit
Deutsch-Jozsa algorithm (a). All beam splitters are 50/50. With beam
splitters C1 and C2 in place, the standard algorithm (b) is performed. In
this work, we show that an alternate encoding (c) is preferable in the
presence of random noise as indicated on rails 2 and 3; replacing beam
splitters C1 and C2 with beam splitters D1 and D2 implements this modified
algorithm.

Fig 2. Experimental setup. Variable-reflectivity beam splitters are
implemented using a pair of polarizing beam splitters (PBS) and a half
waveplate. The ``preparation'' portion of the interferometer produces the
same superposition as the pair of Hadamards in Fig. 1. The oracle consists
of two variable beam splitters which can each be set to swap two rails or
leave them unchanged, plus two half-waveplate used to induce $\pi $ phase
shifts on two of \ the outputs. The random noise is generated by inducing
turbulent airflow under rails 2 and 3 while they are spatially superposed.

Fig 3. Experimental data: Normalized intensity is a measure of the fraction
of photons reaching each detector, PD1 through PD4. Data are shown for both
the DFS and standard encodings, for each of the four oracles (00, 01, 10,
and 11); C indicates ``constant'' oracles while B indicates ``balanced''
oracles. The bottom plot shows the same data in the presence of noise. Note
that the noise has a much more significant effect in the case of the
standard encoding.

Fig 4. The probability of the algorithm returning a 0 or a 1 for each of the
oracles, in each encoding, with and without the addition of phase noise. The
data are extracted by summing the normalized intensities from Fig. 3 for PD1
and PD2 (dashed line, indicating constant oracles) and for PD3 and PD4
(solid line, indicating balanced oracles). Note that the success rate is
close to 1 even in the presence of noise when the DFS encoding is used.

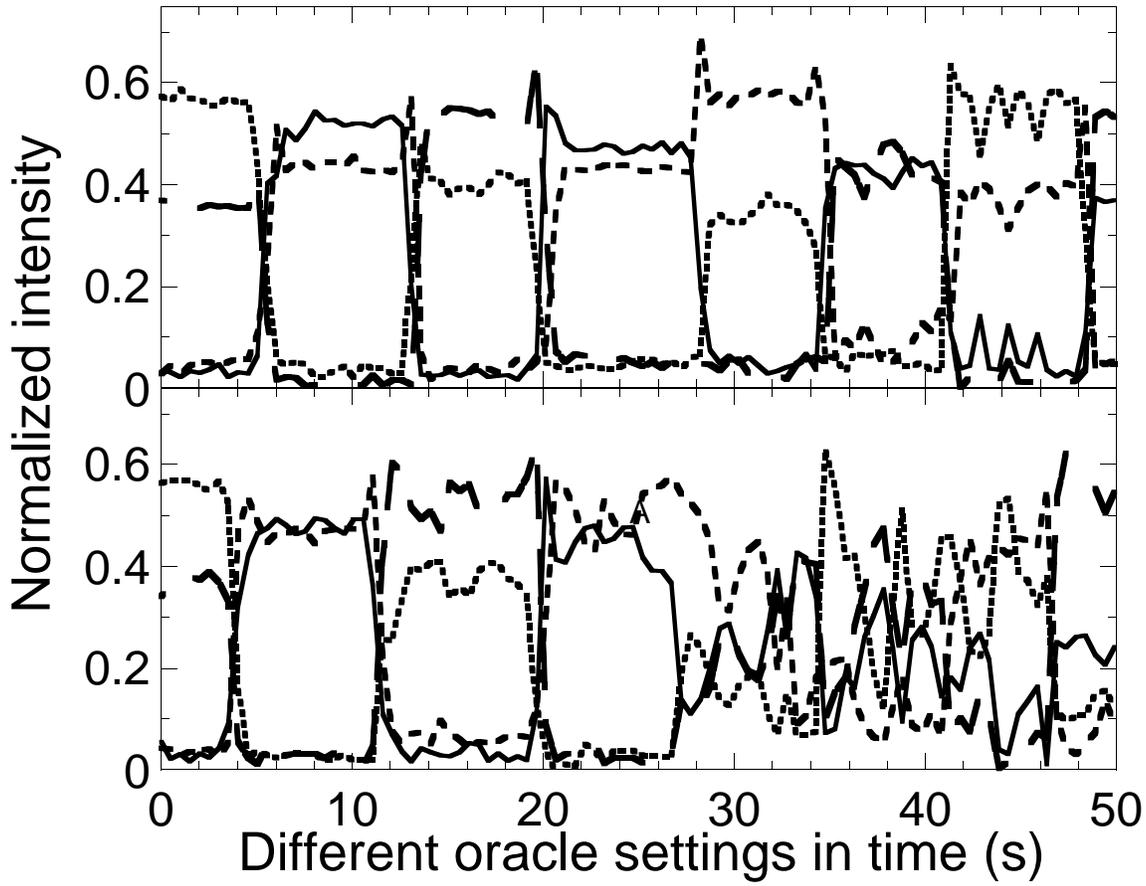
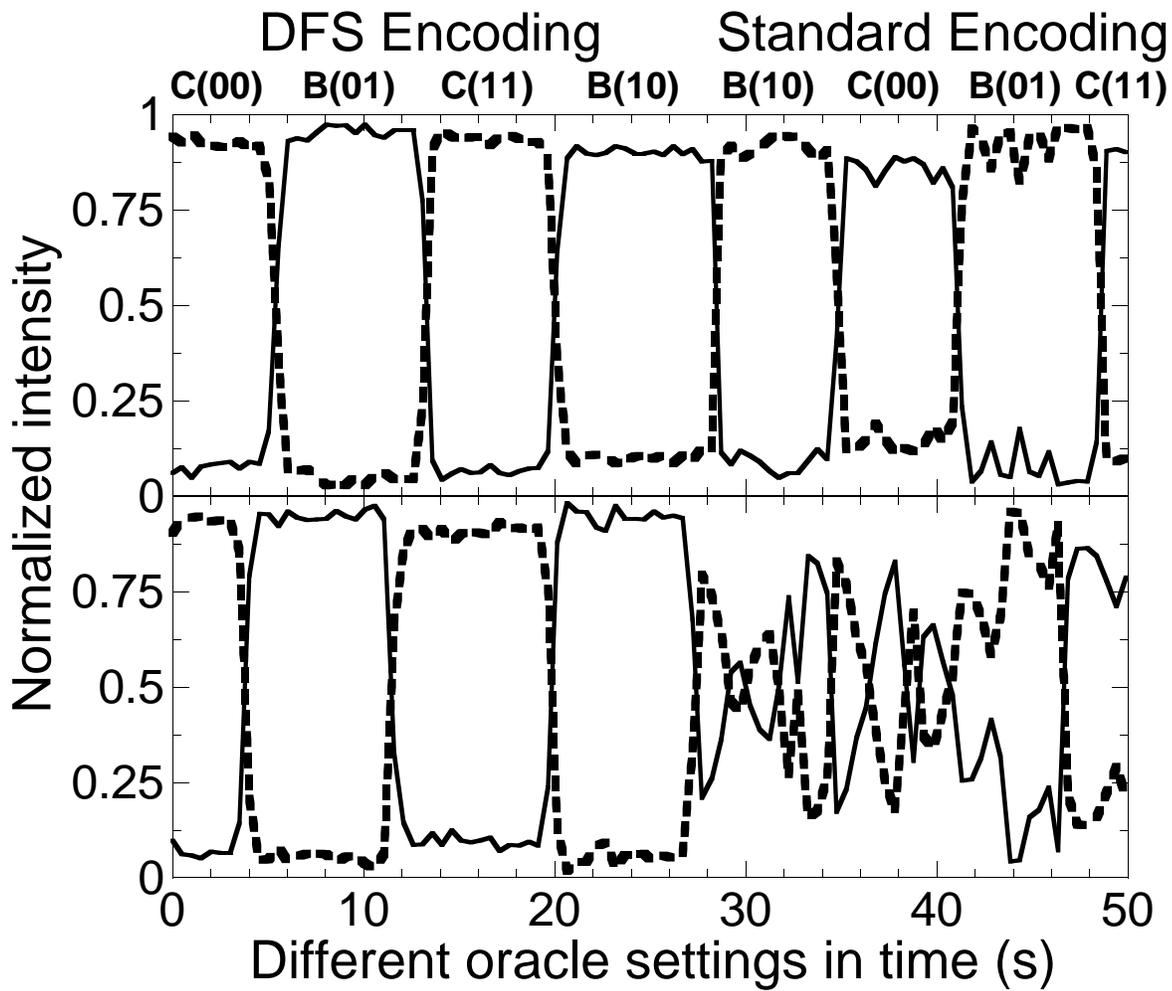
\end{document}